\documentclass[prd,aps,amsmath,nofootinbib,superscriptaddress,showpacs]
{revtex4}

\usepackage{graphicx}
\usepackage{hyperref}
\usepackage{bm}
\usepackage{epsfig}
\usepackage{amsfonts}
\usepackage{mathrsfs}
\usepackage{amsmath}

\catcode`@=11
\let\savesort=\NAT@sort@cites
\newcommand\nosort[1]{\edef\NAT@cite@list{#1}}
\def\citenosort#1{\let\NAT@sort@cites=\nosort \cite{#1}%
   \let\NAT@sort@cites=\savesort}
\catcode`@=12 
\makeatletter
    
    \newcommand{\Rmnum}[1]{\expandafter\@slowromancap\romannumeral #1@}
  \makeatother

\newcommand{\beq}{\begin{equation}}
\newcommand{\eeq}{\end{equation}}

\newcommand{\bea}{\begin{eqnarray}}
\newcommand{\eea}{\end{eqnarray}}
\newcommand{\pa}{\partial}
\newcommand{\bib}{\bibitem}
\newcommand{\comment}[1]{}

\def\hs{\hspace}

\begin{document}

\title{Singularity problem in f(R) model with non-minimal coupling}

\author{Qiang Xu}
\email{xuqiangxu@pku.edu.cn}
\affiliation{Department of Physics, and State Key Laboratory of Nuclear Physics and Technology,\\
Peking University, Beijing 100871, P.R. China}

\author{Sheng-yu Tan}
\email{hilbertan@gmail.com}
\affiliation{Department of Physics, and State Key Laboratory of Nuclear Physics and Technology,\\
Peking University, Beijing 100871, P.R. China}


\begin{abstract}
We consider the non-minimal coupling between matter and the geometry in the f(R) theory. In the new theory which we established, a new scalar $\psi$ has been defined and we give it a certain stability condition. We intend to take a closer look at the dark energy oscillating behavior in the de-Sitter universe and the matter era, from which we derive the oscillating frequency, and the oscillating condition. More importantly, we present the condition of coupling form that the singularity can be solved. We discuss several specific coupling forms, and find logarithmic coupling with an oscillating period $\Delta T\sim\Delta z$ in the matter era $z>4$, can improve singularity in the early universe. The result of numerical calculation verifies our theoretic calculation about the oscillating frequency. Considering two toy models, we find the cosmic evolution in the coupling model is nearly the same as that in the normal f(R) theory when $lna>4$. We also discuss the local tests of the non-minimal coupling f(R) model, and show the constraint on the coupling form.
\end{abstract}

\noindent 

\maketitle

\section{Introduction}
\label{sec:introduction}
The rapid development of observational cosmology starting from 1990s shows that the  expansion of our universe in the present epoch is accelerating. The most popular theory to explain the accelerating universe is dark energy (DE).
One possible dark energy candidate accepted by most of researchers is the cosmological constant. However,
the cosmological constant that remains a mystery, can not be explained clearly in any known theory. If the
cosmological constant originates from the vacuum energy in quantum field theory, as many people believe,
its energy scale is too large to be compatible with the observed dark energy density \cite{la}.
Moreover, the observation indicates that the dark energy equation of the state may cross the phantom divide line $\omega=-1$. This suggests that the cosmological constant \cite{Hannestad:2002ur,Cepa:2004bc,cmb} may not be the only candidate for dark energy. There also exist some other dark energy models, ranging from quintessence, phantom, quintom to chaplygin gas models. For the nice reviews on dark energy, see \cite{Li:2011sd,Copeland:2006wr,Cai:2009zp}.
An alternative scenario for dark energy is infrared(IR) modified gravity.
Among many IR modified gravity models, $f(R)$ gravity is of particular interest.  One important feature in $f(R)$ gravity is the intrinsic existence of an extra dynamical scalar degree of freedom, besides the massless graviton. Therefore it is possible  to study both the early-time inflation and late-time acceleration of the universe in the framework of $f(R)$ gravity, without introducing a scalar field by hand. More interestingly, the effective equation of state could be smaller than $-1$ in $f(R)$ dark energy models, indicating the scalar behaves like a phantom in the Jordan frame.
 The model with a Lagrangian density $f(R)=R-\alpha /R^n$ ($\alpha > 0, n > 0$) was proposed for dark energy \cite{r-n1,ca0,r-n3,Pi:2009an}. However this model is plagued by  matter instability \cite{rn4,rn5} and difficulty to match local gravity constraints. Later on,
researchers have proposed many viable models, seeing \cite{st,hu,tsu,linder,Elizalde:2011ds,Elizalde:2010ts,Cognola:2007zu,Bamba:2010ws,qx}. For nice reviews on f(R) theories, see \cite{Nojiri:2010wj,tsu3}.

Nonetheless, it has been pointed out in Refs. \cite{Elizalde:2011ds,Appleby:2008tv,Tsujikawa:2007xu,Arbuzova:2010iu,Bamba:2011sm,Nojiri:2008fk,Bamba:2008ut} that viable f(R) models generally suffer from a singularity problem. For most of the f(R) models, in order to evade the local gravity tests, $Rf''\ll1$ is a common feature. This leads that the oscillating frequency $F(z)\sim1/\sqrt{Rf''}$ becomes very large. In this case, the Ricci scalar, Hubble parameter and EOS parameter of DE have intense oscillations in the high redshift region. One possible way proposed by \cite{Lee:2012dk} to solve this problem is adding a $R^2/M^2$ term to the f(R) function. However, according to the constraint of f(R) inflation, $M\sim 10^{-6}M_{pl}$. If we consider $R\sim\Lambda$, then the adding term contributes $\Lambda/M^2_{pl}\sim10^{-121}$ to the oscillating frequency. So, this term may has some contribution in high red-shift region such as $z>10^{40}$, but in the matter era $z>4$, such term is of
little help. Even recently, this problem is till be discussed in \cite{odin2012}.

A generalization of the f(R) theories was proposed in \cite{allemandi,inagaki,bert} firstly by including the theory an explicit function of the Ricci scalar R with the matter Lagrangian density $L_m$. As a result of the coupling, shown in \cite{Bertolami:2007gv}, the equation of motion which is the non-geodesic, and an extra force orthogonal to the four-velocity, arise. The implications of the non-minimal coupling on the stellar equilibrium were discussed in \cite{Bertolami:2007vu}, where the constraints on the coupling were obtained. The equivalence between a scalar theory and the model with the non-minimal coupling was considered in \cite{Bertolami:2008zh,Bertolami:2008im}, where the authors showed that the non-minimal coupling f(R) theory corresponds to a two-field scalar theory. Especially in the non-minimal coupling model, the matter part was extended to a arbitrary function of the Lagrangian density of the matter in \cite{Harko:2008qz}. Later, the coupling model was analyzed in \cite{Nesseris:2008mq} to study the matter perturbation and gave a possible way to prevent the f(R) theory conflicting with the galaxies matter spectrum tests, in \cite{Paramos:2011rw} to study the accelerated expansion of the universe, and in \cite{Bertolami:2010ke} to discuss the reheating, which gives a constraint $\Lambda/R_1<10^{-104}$ about the coupling form $f_2=1+(R/R_1)^n$.

In this work, we study the oscillating behavior of the dark energy in the non-minimal coupling f(R) theory. We consider the non-pressure dust as the main contribution of the matter Lagrangian and the coupling form is arbitrary function of Ricci scalar. So the energy-momentum tensor of the matter is generally conserved and the matter density is proportional to $1/a^3$. Similar to the f(R) theories, the non-minimal coupling f(R) model also has a scalar freedom
$\psi$ which consists of the f(R) part $f'_1$, and the coupling part, $\rho f'_2$. Through adding the coupling part, the oscillating frequency $F(z)\sim1/\sqrt{Rf''}$ can be modified, so that the singularity problem may be solved.

This paper is organized as follows. In Section II, we present the definitions in the non-minimal coupling f(R) model and derive the equation of motion. Then we study the stability condition. In Section III and IV, we solve the EOM of dark energy in de-Sitter universe and the matter era theoretically. And we mainly focus on the study of oscillating behavior of the EOS parameter of dark energy. In Section V, we show that a logarithmic coupling can solve the singularity problem. In Section VI, we give the numerical result, and discuss some toy models. In Section VI, we discuss the local tests of the non-minimal coupling f(R) model. Finally, in section VIII, we make the conclusions.

\section{definitions and the stability condition}
\subsection{definitions and the conservation equations}
The action \cite{allemandi} we consider is
\bea
\label{action}
s= \int d^4x\sqrt{-g} \left[\frac{1}{2\kappa^2}f_1(R)+f_2(R)L\right].
\eea
The parameter $\kappa^2=8\pi G$, and we set it to be unit in the following sections. L denotes matter Lagrangian. By varying the action with respect to the metric $g_{\mu\nu}$, following \cite{Bertolami:2007gv,Harko:2008qz,Nesseris:2008mq}with some modifications, we get the modified Einstein equation
\bea
\label{Einst}
(F_1+2LF_2)R_{\mu\nu}-\frac{1}{2}f_1g_{\mu\nu}+(g_{\mu\nu}\Box-\nabla_{\mu}\nabla_{\nu})(F_1+2LF_2)=f_2T_{\mu\nu},
\eea
where $F'_i(R)=f'_i(R)$. The matter energy-momentum tensor is defined as
\bea
\label{et}
T_{\mu\nu}=-\frac{2}{\sqrt{-g}}\frac{\delta(\sqrt{-g}L)}{\delta(g^{\mu\nu})}.
\eea
Using the Bianchi identity, $\nabla^{\mu}G_{\mu\nu}=0$, and the identity
\bea
\label{identity1}
(\Box\nabla_{\nu}-\nabla_{\nu}\Box)F_i=R_{\mu\nu}\nabla^{\mu}F_i,
\eea
and following \cite{Bertolami:2007gv,Harko:2008qz,Nesseris:2008mq,Koivisto:2005yk}, we deduce the following covariant conservation equation
\bea
\label{cons1}
\nabla^{\mu}T_{\mu\nu}=\frac{F_2}{f_2}(g_{\mu\nu}L-T_{\mu\nu})\nabla^{\mu}R,
\eea
which indicates the non-minimal coupling between curvature and matter yields a exchange between matter and the geometry. In the absence of the coupling, $f_2(R)=1$, one recovers the covariant conservation of the energy-momentum
tensor. A simple choice of the matter Lagrangian is $L=-\rho$, if we consider the non-pressure dust as the main contribution. Considering the FRW metric, $T^{00}=\rho$ and $g^{00}=-1$, we find the matter density is conserved
\bea
\label{cons2}
\nabla_0T^{00}=\frac{F_2}{f_2}(g^{00}L-T^{00})=0,
\eea
which yields the conserved equation of the matter density
\bea
\label{cons3}
\dot{\rho}+3H\rho=0.
\eea
So, the evolution of the matter density is the same as that in $\Lambda$CDM, $\rho=\rho_0a^{-3}$.

In a flat FRW metric with a scale factor $a$, according to Eq. (\ref{Einst}), we get the modified Friedmann equations,
\bea
\label{mdfried}
3\psi H^2&=&f_2\rho+\frac{1}{2}(\psi R-f_1)-3H\frac{d}{dt}\psi,\\
\label{mdfried1}
-2\psi \dot{H}&=&f_2\rho+\left(\frac{d^2}{dt^2}-H\frac{d}{dt}\right)\psi,
\eea
where $\psi=F_1+2LF_2$, $H=\frac{\dot{a}}{a}$, $R=12H^2+6\dot{H}$ and the dot denotes a derivative with respect to the cosmic time $t$.

By rewriting Eq. (\ref{mdfried}), we can define the effective energy density,
\bea
\rho_{eff}=3H^2=\frac{1}{\psi} \left\{ f_2\rho+\frac{1}{2}(\psi R-f_1)-3H\frac{d}{dt}\psi\right\},
\eea
and the effective pressure,
\bea
P_{eff}=-(2\dot{H}+3H^2)=\frac{1}{\psi} \left\{-\frac{1}{2}(\psi R-f_1) +\left(\frac{d^2}{dt^2}+2H\frac{d}{dt}\right)\psi\right\}.
\eea
We can define the dark energy density $\rho_{de}$ as $\rho_{de}=\rho_{eff}-\rho$, and we use the definition in the
previous papers \cite{hu,Elizalde:2011ds},
\bea
\label{dedefine}
y(z)\equiv\frac{\rho_{de}}{\widetilde{m}^2}=\frac{H^2}{\widetilde{m}^2}-(1+z)^3.
\eea
Here $\widetilde{m}^2$ is the mass scale $\widetilde{m}^2=\kappa^2\rho_0/3$ and $\rho_0$ is the matter density at the present time.

The EOS-parameter for dark energy can be expressed by $y(z)$ as
\bea
\label{wde}
\omega_{de}=-1+\frac{1}{3}(1+z)\frac{1}{y(z)}\frac{dy(z)}{dz}.
\eea
When $y(z)$ tends to be a constant, the EOS-parameter tends to be -1.

\subsection{field equations and the stability condition}
In this section, we discuss the stability in the local gravity.
Taking the trace of Eq. (\ref{Einst}), we get
\beq
\label{scalar}
3\Box\psi+\psi R-2f_1(R)-f_2T=0.
\eeq
First, we decompose the quantities R, $\psi(R)$ and $T_{\mu\nu}$ into the background part with a constant curvature and the perturbed part:
$R=R_0+\delta R$, $\psi=\psi_0+\delta\psi$ and $T=T_0+\delta T$. We consider R close to the mean-field value $R_0$, and the metric still very close to the minkowski case. The linear expansion of Eq. (\ref{scalar}) in a time-independent background gives
\beq
\label{de}
\nabla^2\delta\psi-M_{\psi}^2\delta\psi=\frac{f_2}{3}\delta T,
\eeq
where $M_{\psi}$ is the mass of the scalar
\bea
\label{mass}
M^2_{\psi}=\frac{R}{3}\left(\frac{F_1-F_2T_0}{R\psi'}-1\right),
\eea
and $\psi'=F'_1+2T_0F'_2$. The stability condition is given by
\bea
\label{stable}
0<\frac{R\psi'}{F_1-F_2T_0}<1.
\eea
Note that, to respect the solar system constraints, we must let $f_2\rightarrow1$, $F_1\rightarrow1$ and $F_2\rightarrow0$. So this condition (\ref{stable}) can be rewritten approximately as
\bea
0<R\psi'<1.
\eea
Especially in the late time de-Sitter universe, $\rho=-T_0\rightarrow0$, $\psi\rightarrow F_1$ and $\psi'\rightarrow F'_1$, then the stability condition (\ref{stable}) recovers to that in the normal f(R) gravity.

\section{Oscillations in the de-Sitter universe}
The trace equation (\ref{scalar}) of the field can be recast in the form
\beq
\label{veff}
\Box\psi=\frac{\pa V_{eff}}{\pa\psi}.
\eeq
Similar to the dynamic of normal f(R) gravity, there is a new scalar freedom $\psi$ which decides the cosmic evolution, and the effective potential has a minimum at
\beq
\label{de}
\frac{\pa V_{eff}}{\pa\psi}=-\frac{1}{3}\left[R\psi-2f_1-f_2T\right]=0.
\eeq
In the de-Sitter universe, neglecting the contribution of the matter, the effective potential has a minimum at
\bea
\label{desolut}
R\psi-2f_1=0,
\eea
where $R_{ds}$ is a constant. Using Eq. (\ref{dedefine}), the Ricci scalar can be expressed as
\bea
\label{ricci}
R=3\widetilde{m}^2\left(4y(z)-(1+z)\frac{dy(z)}{dz}+(1+z)^3\right),
\eea
By combining Eq. (\ref{mdfried}) with Eq. (\ref{ricci}), one gets
\bea
\label{eom}
\frac{d^2y(z)}{dz^2}+\frac{a}{1+z}\frac{dy(z)}{dz}+\frac{b}{(1+z)^2}y(z)=c(1+z)+\frac{d}{(1+z)^2},
\eea
where
\bea
a&=&-3-\frac{1-\psi}{6H^2\psi'},\\
b&=&\frac{2-\psi}{3H^2\psi'},\\
c&=&3+\frac{1-\psi}{6H^2\psi'}+\frac{6F_2}{\psi'},\\
d&=&\frac{R-f_1}{6H^23\widetilde{m}^2\psi'},
\eea
and we have used $d/dt=-(1+z)H(z)d/dz$.

When $z\rightarrow-1^{+}$, we consider the perturbations around the de-Sitter solution of the dark energy density
$y(z)\rightarrow y_0$, hence
\bea
R\rightarrow R_{ds}\simeq12\widetilde{m}^2y_0\simeq12H_{ds}^2,\\
\rho=3\widetilde{m}^2(1+z)^3\rightarrow0,\\
\psi=F_1-2\rho F_2\rightarrow F_1, \psi'\rightarrow F'_1.
\eea
In this case, we can rewrite the coefficients as
\bea
a&=&-3-\frac{2(1-\psi)}{R\psi'},\\
b&=&\frac{4(2-\psi)}{R\psi'},\\
c&=&3+\frac{2(1-\psi)}{R\psi'}+\frac{6F_2}{\psi'},\\
d&=&\frac{2(R-f_1)}{R3\widetilde{m}^2\psi'}.
\eea

The solution of Eq.(\ref{eom}) is
\bea
\label{solu1}
y(z)=y_0+B(1+z)^{\frac{1}{2}(1-a\pm\sqrt{(a-1)^2-4b})}+\frac{c}{3a+b+6}(1+z)^3,
\eea
where B is a constant depending on the initial condition and $y_0=d/b$. Depending on the sign of the discriminant in the square root of Eq. (\ref{solu1}), there are two possible behaviors for this model. If $(a-1)^2-4b>0$, the solution approaches the de-Sitter point as a power function of (1+z). Otherwise if $(a-1)^2-4b<0$, the dark energy shows an oscillating behavior as
\bea
\label{osc}
y=y_0+2B(1+z)^{\frac{1-a}{2}}\cos(\frac{\sqrt{(a-1)^2-4b}}{2}\ln(1+z))+\frac{c}{3a+b+6}(1+z)^3.
\eea
Now we write out the discriminant,
\bea
\label{disciminant}
(a-1)^2-4b=4\frac{\psi}{R\psi'}-\left(2-\frac{1-\psi}{R\psi'}\right).
\eea
Combining the stability condition Eq. (\ref{stable}), $1-\psi>0$ and $R\psi'>0$ we know,
when
\bea
\label{cd1}
\frac{R\psi'}{\psi}<\frac{1}{4}\left(2-\frac{1-\psi}{R\psi'}\right),
\eea
the dark energy has an oscillating behavior near the de-Sitter point. In this case, according to the definition in Eq. (\ref{wde}), one has
\bea
\omega_{de}=-1+\frac{4Bb}{3y_0}(1+z)^{\frac{1-a}{2}}\cos\left[\frac{\sqrt{(a-1)^2-4b}}{2}\ln(1+z)+\alpha\right]
+\frac{c}{y_0(3a+b+6)}(1+z)^3,
\eea
where $\alpha=\arctan \sqrt{(a-1)^2-4b}/(1-a)$.
The EOS parameter of dark energy also has an oscillating form. We have noticed that, on condition $1-a>0$, the oscillating amplitude becomes smaller and smaller. Therefore, as time goes on, $\omega_{de}$ tends to be -1. If we take the e-folding number $N=-\ln (1+z)$ as the variable, the oscillating frequency is
\bea
F(z)=\sqrt{\frac{\psi}{R\psi'}-\frac{1}{2}\left(2-\frac{1-\psi}{R\psi'}\right)^2}.
\eea

\section{Oscillating in the matter era}
In the matter dominated era, $z>3$ and $R=3H^2$, in order to match the local gravity tests, the following conditions
\bea
\label{codition}
f_1\rightarrow R-2\Lambda, F_1\rightarrow1,\\\nonumber
f_2\rightarrow1, F_2\rightarrow0, \psi\rightarrow1, \psi'\rightarrow0,
\eea
must be satisfied. And the minimum point of the effective potential in the Eq. (\ref{veff}) is
\bea
R\psi-2f_1-f_2T=0.
\eea
So, we neglect the dark energy contribution, and we write out the expression of Ricci scalar
\bea
R\simeq\rho=3\widetilde{m}^2(1+z)^3.
\eea
In this case, according to Eq. (\ref{eom}), the EOM in the matter era is
\bea
\label{eom2}
\frac{d^2y(z)}{dz^2}+\frac{-3-\frac{1-\psi}{2R\psi'}}{1+z}\frac{dy(z)}{dz}+\frac{(2-\psi)}{R\psi'(1+z)^2}y(z)\\\nonumber
=\left(3+\frac{1-\psi}{2R\psi'}+\frac{6F_2}{\psi'}+\frac{1-f_1/R}{2R\psi'}\right)(1+z).
\eea
Using the method in the literature \cite{Elizalde:2011ds}, with a little modification, we solve the EOM near the minimum of the effective
potential $z=z_0+(z-z_0)$, and $z-z_0\ll z_0$. To first order in $(z-z_0)$, the Eq. (\ref{eom2}) changes into
\bea
\label{eom3}
\frac{d^2y(z)}{dz^2}+\frac{a}{1+z_0}\frac{dy(z)}{dz}+\frac{b}{(1+z_0)^2}y(z_0)=c(z-z_0)+d(1+z_0),
\eea
where
\bea
&&a=-3-\frac{1-\psi(R_0)}{2R_0\psi'(R_0)},\\
&&b=\frac{2-\psi(R_0)}{R_0\psi'(R_0)},\\
&&c=\frac{3}{2}+\frac{-4-\psi(R_0)+5f_1(R_0)/R_0}{2R_0\psi'(R_0)}-
\frac{3\psi''(R_0)(2-\psi(R_0)-f_1(R_0)/R_0)}{2\psi'^2(R_0)}\\\nonumber
&&+\frac{6(F_2(R_0)+3R_0f''_2(R_0))\psi'(R_0)-6F_2(R_0) 3R_0(\psi''(R_0)-2f''_2(R_0))}{\psi'^2(R_0)},\\
&&d=3+\frac{1-\psi(R_0)}{2R_0\psi'(R_0)}+\frac{6F_2(R_0)}{\psi'(R_0)}+\frac{1-f_1(R_0)/R_0}{2R_0\psi'(R_0)}.
\eea
The solution of Eq. (\ref{eom3}) is
\bea
\label{solu2}
y(z)=\frac{d}{b}(1+z_0)^3+\frac{c(1+z_0)^2}{b}(z-z_0)-\frac{ac}{b^2}(1+z_0)^3
+C_0e^{\frac{-a\pm\sqrt{a^2-4b}}{2(1+z_0)}(z-z_0)},
\eea
where $C_0$ is a constant depending on initial conditions. Generally speaking, for different forms of $\psi$ , $1/R^n$ for example, corresponds to $(1-\psi)/R\psi'\sim\mathcal{O}(1)$ and $e^{-R}$ corresponds to $(1-\psi)/R\psi'\sim0$. Note that $R\psi'\ll1$. Therefore, $a\sim\mathcal{O}(1)$, $b\sim1/(R\psi')\gg1$, $c\sim d\sim(\Lambda/R)/(R\psi')$ and $ac/b^2\ll1$ which mean we can neglect the third term in Eq. (\ref{solu2}). In this case, the discriminant in the square root of Eq. (\ref{solu2}) must be negative, and the solution (\ref{solu2}) becomes an oscillating form
\bea
y(z)&\simeq y_0+\frac{c}{b}(1+z_0)^2(z-z_0)
+2C_0e^{\frac{-a}{2(1+z_0)}}\cos\left(\frac{\sqrt{4b-a^2}}{2(1+z_0)}(z-z_0)\right),
\eea
where $y_0=d(1+z_0)^3/b$. Using Eq.(\ref{wde}) again, we give EOS parameter of dark energy near the redshift $z=z_0$,
\bea
\label{wde2}
\omega_{de}=-1+\frac{c}{3by_0}(1+z_0)^3-
C_0\frac{\sqrt{4b-a^2}}{3y_0}e^{\frac{-a}{2(1+z_0)}}\sin\left(\frac{\sqrt{4b-a^2}}{2(1+z_0)}(z-z_0)\right).
\eea
Since we care about the oscillating, we present the oscillating frequency,
\bea
\label{fz1}
F(z_0)\simeq\frac{1}{(1+z_0)\sqrt{R_0\psi'}},
\eea
which corresponds to that in normal f(R) theory \cite{Elizalde:2011ds}, where we have used the condition $a^2\ll4b$.

If we hope the non-minimal coupling f(R) model improve the singularity problem, according to the expression (\ref{fz1}),it is required that $R\psi'(1+z)$ must not be the decreasing function of Ricci scalar. In the matter dominated era, we ignore the normal $f_1(R)$ part. On the other hand, by considering think the scalar field $\psi$ stays at the minimal point of the effective potential, or equivalently $R\simeq\rho$, we get
\bea
R\psi'\simeq-f_2''R^2.
\eea
Recalling the stability condition (\ref{stable}), we get the constraint of $f_2$
\bea
0<-f_2''R^2<1.
\eea
On the other hand, to evade the singularity problem, considering $1+z\sim R^{1/3}$, $-f_2''R^{7/3}$ must not be a decreasing function of Ricci scalar. Therefore we get the condition that improve the singularity,
\bea
\label{panju}
-\frac{f'''_2R}{f''_2}\leq\frac{7}{3}.
\eea
\section{specific couplings}
In the normal f(R) gravity, there is a singularity problem that the oscillating frequency is too large due to $Rf''(R)\ll1$. We  expect to reduce the oscillating frequency by adding a coupling term. Here, we consider some specific models.
\subsection{logarithmic coupling}
Basing on the f(R) theory, we consider a logarithmic coupling
\bea
f_2(R)=1+\alpha\ln(R/c),
\eea
where $c$ is the cosmological constant sharing the same unit with the cosmological constant $\Lambda$, and the parameter $\alpha$ is a small non-dimensional constant. In the matter era, for most of f(R) theory, to evade the solar system tests, the $Rf_1''$ is set to be very small. Here, we set $Rf_1''\ll\alpha\ll1$ reasonably. Such a coupling has a virtue that it has little influence on the evolution of the early universe, for example, inflation and reheating era, as long as we take a proper coupling constant $\alpha$. In the inflation and reheating era, considering c is a cosmological constant $\Lambda$, $R/c\sim10^{121}$, we get $f_2\sim1+\alpha121\ln10$, especially when $\alpha<10^{-4}$, we get $f_2\sim1$ approximately. This recovers the normal f(R) theory. As for the local gravity, $R/c\sim10^5$, however, the detailed discussion, which we put in the Sec. VII, may be more complex. Here, we just set $\alpha$ to be a small constant without dimension.
Therefore, the scalar $\psi$ is
\begin{equation}\label{psi}
 \psi=f_1'-2\alpha\frac{\rho}{R}
\end{equation}
where we have used the approximation $R\sim\rho$. According to the definition, $\psi'$ can be rewritten as
\begin{equation}\label{psi1}
  \psi'=f_1''+2\frac{\alpha\rho}{R^2}\simeq f_1''+2\frac{\alpha}{R}\simeq2\frac{\alpha}{R}.
\end{equation}
So, we have the approximation $\psi\simeq1-2\alpha$ and $R\psi'\simeq2\alpha$.
In this case, we can get the oscillating frequency in the matter era,
\begin{equation}\label{fz}
  F(z_0)=\frac{1}{(1+z_0)\sqrt{2\alpha}}=\frac{2\pi}{T},
\end{equation}
which indicates that, the period $T$ is proportional to the red-shift z if we consider a small logarithmic coupling. This solves the singularity problem in the matter era.
\subsection{power-law coupling}
We also consider the power-law coupling, assuming that
\bea
f_2(R)=1+\left(\frac{R}{R_n}\right)^n,
\eea
where $R_n$ is a constant with the same unit as $M^2_{pl}$.

Firstly, we discuss the stability condition in power-law coupling, and we get
\bea
R\psi'=f''_1R-2\rho R \frac{n(n-1)}{R^2_n}\left(\frac{R}{R_n}\right)^{n-2}\simeq=f''_1R-2n(n-1)\left(\frac{R}{R_n}\right)^n.
\eea
In order to improve the large oscillating frequency, assuming $(R/R_n)^n\gg f''_1R$ in the matter era,
according to Eq. (\ref{stable}), one has
\bea
0<-2n(n-1)\left(\frac{R}{R_n}\right)^n<1.
\eea
Then we have two condition
\bea
0<n<1,\left(\frac{R}{R_n}\right)<\frac{1}{2n(1-n)}.
\eea
Therefor, we do not consider the inverse power-law coupling $n<0$ which is used to mimic the dark matter in \cite{Paramos:2011rw}, either the case $n\geq2$. Using condition (\ref{panju}), we get $n<13/3$, which is always satisfied, if we consider stability condition.

Now we set $n=1-\epsilon$ in order that $f''_2\neq0$, where $\epsilon\ll1$ is a small positive constant.
Recalling the constraint arising from the non-minimal coupling scenario for reheating\cite{Bertolami:2010ke,Bertolami:2011fz},
\bea
\frac{\Lambda}{R_1}<10^{-104},
\eea
we know, in the matter era, though condition (\ref{stable}) and (\ref{panju}) is satisfied, $(R/R_1)^{1-\epsilon}$ is so small that it has no contribution to $R\psi'$. Hence, the power-law coupling is unavailable to improve the singularity problem.

\subsection{exponential coupling}
Ones assume a power-law exponential coupling of the form
\bea
f_2=e^{\left(\frac{R}{R_n}\right)^n},
\eea
where the $\left(\frac{R}{R_n}\right)^n$ must be very small in the matter era to recover the normal f(R) theory.
We get the $R\psi'$,
\bea
\begin{aligned}
R\psi'&=f''_1R-2\rho R ne^{\left(\frac{R}{R_n}\right)^n} \left(\frac{R}{R_n}\right)^{n-2}\left[(n-1)+n\left(\frac{R}{R_n}\right)^n\right]\\
&\simeq f''_1R-2n(n-1)\left(\frac{R}{R_n}\right)^n.
\end{aligned}
\eea
Therefor the discussion is the same as the power-law coupling.

There are also some other exponential coupling, such as
\bea
f_2=1-e^{-R/c},
\eea
where the c is a constant which sharing the same unit and order with the cosmological constant $\Lambda$.
In this model, $f''_2\sim e^{-R/c}$ decreases rapidly that means $R\psi'$ is still very small in the matter era.
So, the exponential coupling is unavailable to improve the singularity problem either. Using condition (\ref{panju}),
we get
\bea
\label{increa}
-\frac{f'''_2R}{f''_2}=R/c,
\eea
which is bigger than 7/3.

\section{numerical result and the late time evolution}
In this section, using numerical calculation, we examine and certify the theoretical calculation about the oscillating frequency in the matter era. Then, using several possible coupling forms, we discuss the behavior EOS parameter in the de-Sitter universe.
\subsection{non-singularity with a small logarithmic coupling}
We set the coupling constant $\alpha=10^{-10}$, and the initial condition $z_0=8$, $y(z_0)=2\Lambda$, $y'(z_0)=0$ . Therefore, in the matter era, seeing Eq. (\ref{psi}) and Eq. (\ref{psi1}), when the contribution of $f_1$ is insignificant comparing with the coupling term, the coupling constant will dominate. This will suppress the oscillating in the early universe.
Fig. 1 shows the EOS parameter of dark energy as a function of red-shift. In the late time, with a small coupling, we can not distinguish the result from that in normal f(R) theory without coupling. However, in the matter era, there are some differences. Fig. 2 shows the oscillating at different red-shift. Note that, as the red-shift gets bigger, the oscillating period getting bigger. It reads as:\\
{\begin{tabular}{@{}ccccc@{}cccccc@{}}
&& redshift &   \hs{8ex}   &numerical calculating period & \hs{10ex} &theoretic predict\\
&& z=4      &   \hs{8ex}  T=&0.00042  &\hs{10ex} &0.00044\\
&& z=5      &   \hs{8ex}  T=&0.00053  &\hs{10ex}  &0.00053\\
&& z=6      &   \hs{8ex}  T=&0.00062  &\hs{10ex} &0.00062\\
&& z=7      &   \hs{8ex}  T=&0.00071  &\hs{10ex} &0.00071\\
\end{tabular} \label{Table}}

Comparing with the result in the literature\cite{Elizalde:2011ds}, there is no more singularity in this model.
Form the Table. 1, we get the relationship
\bea
\Delta T\sim \Delta z,
\eea
which can be also got from the theoretical result Eq. (\ref{fz}). Note that, when $z=4$, there is some differences between the numerical and the theoretical calculation. This is because our approximation $\psi\rightarrow1$ and $R\psi'\ll1$ is no longer valid in the low redshift region. However when $z>4$, our theoretical calculation is still valid.

\begin{figure}
\label{os2}
\psfig{figure=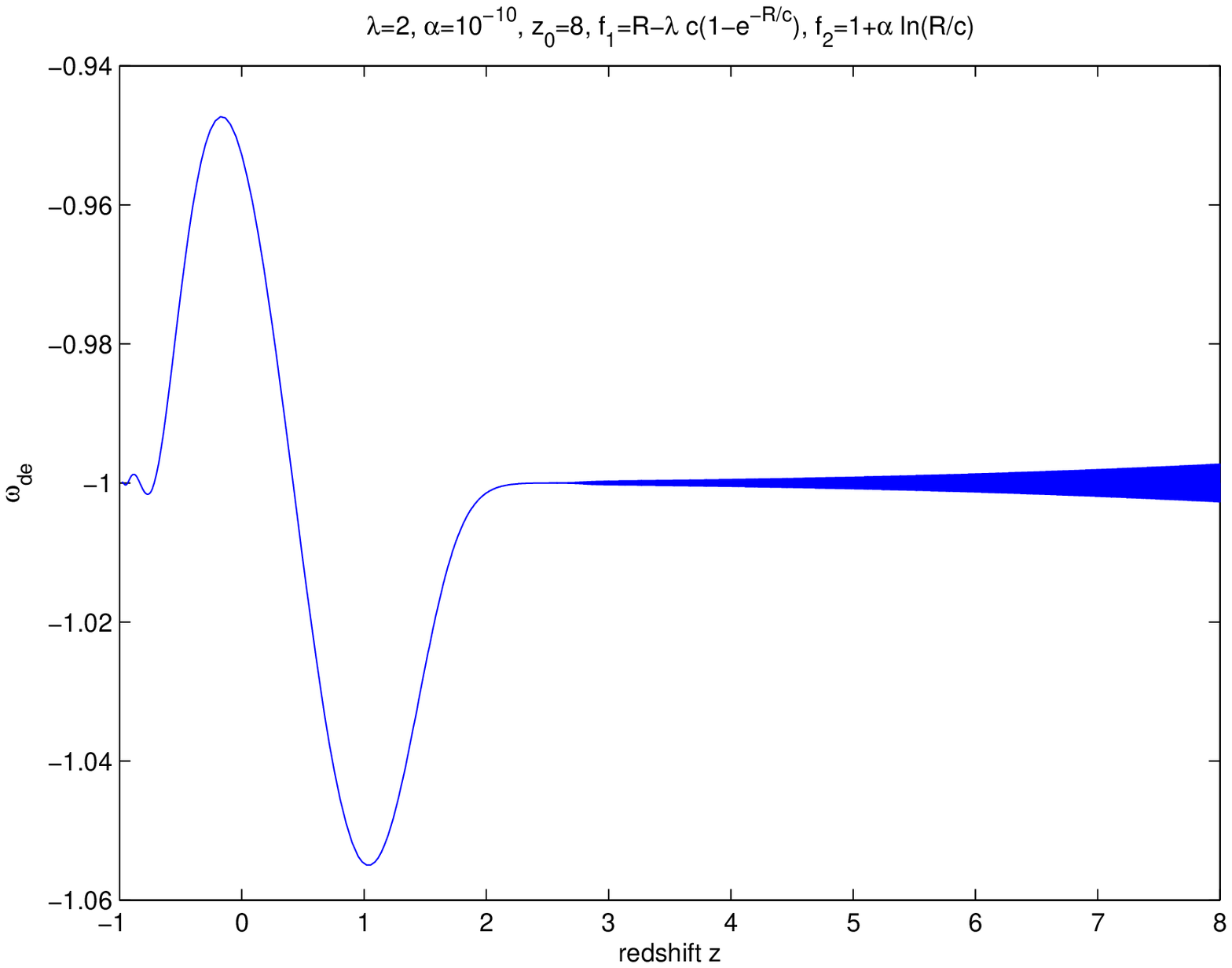,width=10cm,height=8cm}
\caption{}
\end{figure}

\begin{figure}
\label{os1}
\psfig{figure=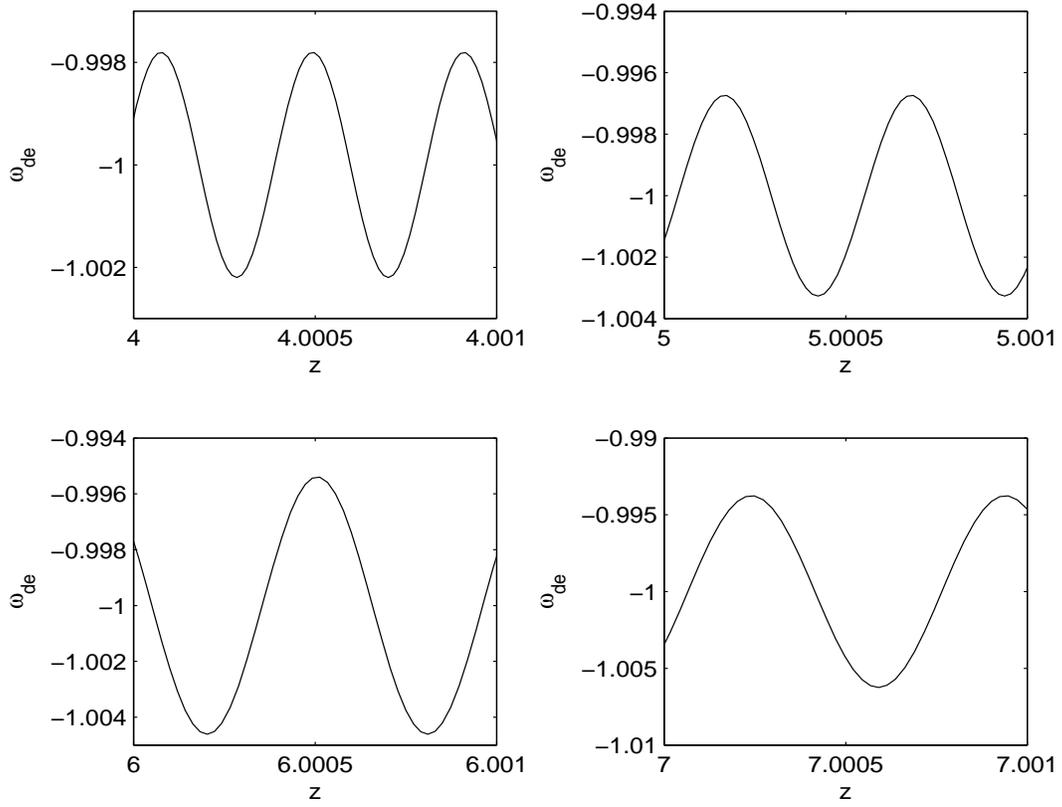,width=16cm,height=12cm}
\caption{The EOS parameter of dark energy at different red-shift, }
\end{figure}

\subsection{late time evolution in toy models}
If we ignore the singularity problem in the early universe and care more about the late time universe, we want to check how the non-minimal coupling wound influence the late time universe. To match the solar system tests, when Ricci scalar is large, $f_2(R)$ must tend to be 1 while considering the coupling form. We still consider a exponential model, and with a exponential coupling:
\bea
f_1(R)&=&R-\lambda c(1-e^{-R/c}),\\
f_2(R)&=&1-e^{-R/c}.
\eea
When $\lambda>1$, in the matter era, $e^{-R/c}\ll1$ which assure $f_2\rightarrow1$.
Fig. 3 shows the detail. when $-1<N<3$, we can distinguish the results from that in normal f(R) theory. When $N>3$,
the result is nearly the same. In this model, we set $\lambda=2$ and the initial red-shift is $z_0=3$.
\begin{figure}
\psfig{figure=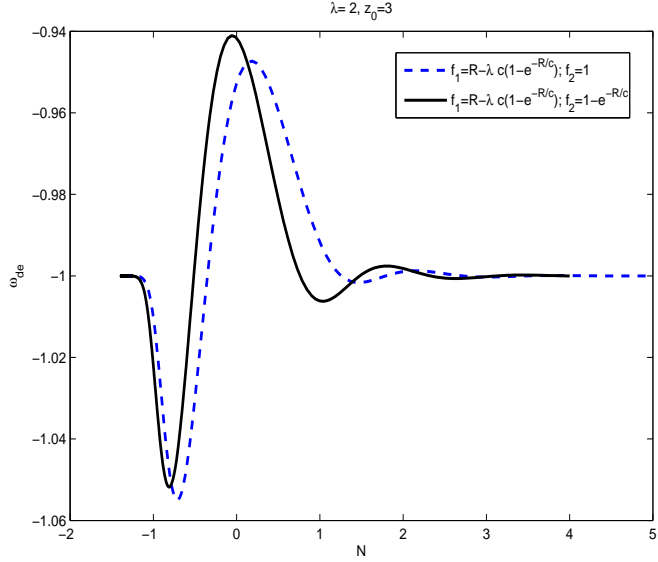,width=10cm,height=8cm}
\caption{The EOS parameter of dark energy in exponential model with a exponential coupling, where the parameter $\lambda=2$}
\end{figure}

We also consider the model
\bea
f_1(R)=(R-\lambda c)e^{\lambda(\frac{c}{R})^n}
\eea
 proposed in \cite{qx} recently. Therefore we add new exponential coupling form
\bea
f_2(R)=e^{\lambda_1(\frac{c}{R})^n}
\eea
when $\lambda>1$, in the matter era, $(c/R)^n\rightarrow0$ that assure $f_2\rightarrow1$.
where the parameters we set is $\lambda=1$, $n=4$ and $\lambda_1=0.5$. the initial red-shift is $z_0=12$.
\begin{figure}
\psfig{figure=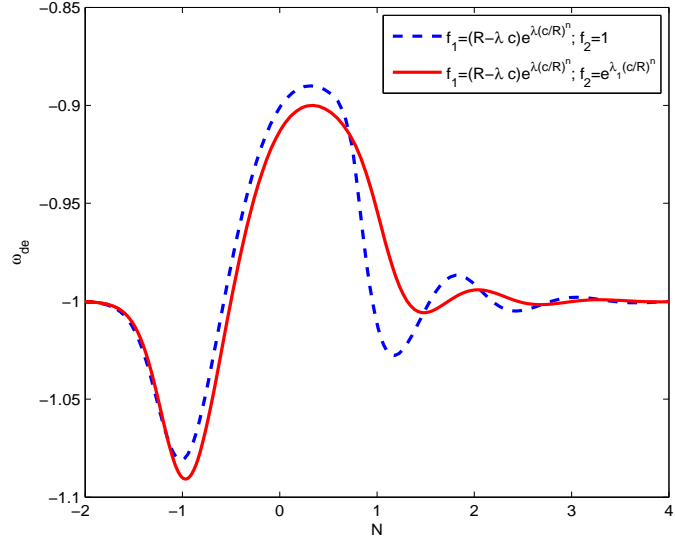,width=10cm,height=8cm}
\caption{The EOS parameter of dark energy in new exponential model with a new exponential coupling}
\end{figure}
Fig. 4 shows the details. In this coupling model, the oscillating amplitude is reduced comparing with the normal f(R)
theory.

\section{local tests}
In this section, we consider local tests of the non-minimal coupling f(R) model. We will use the similar method by Hu and Sawicki \cite{hu}. We consider a spherically symmetric isotropic metric,
\bea
ds^2=-\left[1-2\mathcal{A}(r)+2\mathcal{B}(r)\right]dt^2+\left[1+2\mathcal{A}(r)\right](dr^2+r^2d\Omega),
\eea
where we assume $\mathcal{A}\ll1$ and $\mathcal{B}\ll1$ near the source. especially, in the GR limit, $\mathcal{B}\rightarrow0$ since the solar system tests provide the strongest limits on $\mathcal{B}$.

Generally, according to the definition of the Ricci tensor, we get
\bea
R^0_0=\nabla^2(\mathcal{A-B}),\\
R=-2\nabla^2(\mathcal{A+B}).
\eea
And form the modified Einstein equation (\ref{Einst}), we get the time-time component of the field equation
\bea
\psi R^0_0-\frac{1}{2}f_1(R)+\Box\psi+\pa^2_t\psi=-f_2\kappa^2\rho.
\eea
If we consider the static solution, combining this equation with the trace equation (\ref{scalar}), we get the time-time component of Ricci tensor
\bea
3R^0_0=-\frac{\frac{1}{2}f_1+2f_2\kappa^2\rho}{\psi}+R.
\eea
Then we get
\bea
\nabla^2\mathcal{A}=\frac{1}{2}R^0_0-\frac{1}{4}R=-\frac{1}{12}\left(R+\frac{f_1+4f_2\kappa^2\rho}{\psi}\right),\\
\nabla^2\mathcal{B}=-\frac{1}{2}(R^0_0+\frac{R}{2})=\frac{1}{12}\left(-5R+\frac{f_1+4f_2\kappa^2\rho}{\psi}\right).
\eea
In the high curvature region, for general f(R) model, we have $f_1\rightarrow R$ and $f'_1\rightarrow1$. On the other hand, taking proper form of $f_2$, which has the limit $f_2\rightarrow1$ and $f'_2\rightarrow0$ in the high curvature region, we have $\psi\rightarrow1$. So we consider
\bea
\nabla^2\mathcal{A}\approx-\frac{1}{6}-\frac{1}{3}\kappa^2\rho,\\
\nabla^2\mathcal{B}\approx\frac{1}{3}(\kappa^2\rho-R).
\eea
In the same limit, from the trace equation of the field equation (\ref{scalar}), we have
\bea
\nabla^2\psi\approx\frac{1}{3}(R-\kappa^2\rho).
\eea
Therefore, $\psi$ together with $\mathcal{B}$ has a solution
\bea
\psi(R)+\mathcal{B}(r)=a_1+\frac{a_2}{r}.
\eea
We assume the solution remains finite at $r=0$, $a_2=0$, and when $r\rightarrow\infty$ we assume $\mathcal{B}(\infty)\rightarrow0$. So, we have
\bea
\mathcal{B}(r_{\odot})=-[\psi(R_{\odot})-\psi(R_{\infty})]
\eea
It is difficult to find an accurate solution of the Ricci scalar. However, we can consider the field stays at the minimal point $R^{(0)}=\kappa^2\rho$ approximately, and the deviation from the minimum is
\bea
R^{(1)}=3\Box\psi(R)+(\psi-1)R-2(f_1-R)+(f_2-1)\kappa^2\rho,
\eea
where the last three terms we know is small, so we care the field gradients $\nabla^2\psi$. And we define $R=R^{(0)}+R^{(1)}$. A sufficient condition for $R^{(1)}\ll R^{(0)}$ is the compton condition
\bea
\psi'|_{R=\kappa^2\rho}\partial^2_i\rho\ll\rho.
\eea
If this condition is always satisfied, we consider $R\approx\kappa^2\rho$. If this condition is violated beyond some outer radius, such as the sun radius, this approximation is not valid until the compton condition is locally satisfied again outside the sun. Then the the field stays at the minimum again.

Anyway, the curvature near the sun is high, so we think $\psi(R_{\odot})\approx1$ comparing with $\psi(\infty)$.
On the other hand, we use the galaxy radius instead of $\infty$, considering the field stay at the minimum point $R_{\infty}=R_{g}=\kappa^2\rho_g$. So we get
\bea
\mathcal{B}(r_{\odot})\approx\psi(R_{g})-1
\eea

Finally the deviation from the GR metric is given by
\bea
\gamma-1\equiv\frac{\mathcal{B}}{\mathcal{A}-\mathcal{B}}
\eea
where $\mathcal{A}$ we take is the Newtonian potential of the sun $\Phi_N=GM/r=2.12\times10^{-6}$. The tightest experimental bound on the PNP parameter is given by $|\gamma-1|<2.3\times10^{-5}$ \cite{616,617,bb}. Therefore
we get the constraint
\bea
\psi(R_{g})-1=(f'_1(R_g)-1)-2f'_2\rho<0.5\times10^{-10}.
\eea
Generally, $(f'_1(R_g)-1)<0.5\times10^{-10}$ is the constraint of normal f(R) component. If we require
\bea
2f'_2(R_g)\rho_g< 0.5\times10^{-10},
\eea
the coupling model may evade the solar system test. For logarithmic coupling, $f'_2=2\alpha\rho_g/R_g\approx2\alpha$,
the constraint is $\alpha<10^{-10}$.
\section{Conclusions and discussion}
We consider a non-minimal coupling between geometry and matter in the f(R) theory. We find this theory also has a independent scalar $\psi=f_1'+2Lf_2'$ similar to the scalar $f'$ in the normal f(R) theory. We give its stability condition (\ref{stable}), which is also similar to normal f(R) theory.

We discuss the evolution of dark energy in the de-Sitter universe, when the condition (\ref{cd1}) is satisfied, the dark energy density will have oscillating behavior. Meanwhile, we find the dark energy is always oscillating in the matter era, with an oscillating frequency $F(z)\propto1/(1+z)\sqrt{R\psi'}$.  Adding the coupling part into the f(R) theory, under the condition that $-\frac{f'''_2R}{f''_2}<\frac{7}{3}$, the oscillation is suppressed successfully by the coupling term in the matter era.

Three specific coupling forms have been studied, and it turns out that the logarithmic coupling, with a oscillating period $\Delta T\sim\Delta z$, is a good choice. The power law coupling also can suppress the oscillating becoming strong. However, the reheating constraint $\frac{\Lambda}{R_1}<10^{-104}$ gives a overlarge energy scale. And this will disappointedly lead to very small $R/R_1$, which makes it difficult to reduce the oscillating frequency in the matter era. The exponential coupling $f_2=1-e^{-R}$ can not satisfy the increasing function condition (\ref{increa}).

Through numerical calculation, we verified our theoretical calculation of the oscillating frequency in the matter era, and found only when $z\leq4$ the numerical calculation has some differences with the theoretical calculation. This is because our approximation $\psi\rightarrow1$ and $R\psi'\ll1$ is no longer valid in the low redshift region. However when $z>4$, our theoretical calculation is still valid. We take two toy models as examples, and show that the non-minimal coupling does not change the cosmic evolution of f(R) model in the late universe. Especially when $lna>4$, we can not distinguish between coupling model and the f(R) model.

In the end, we discuss the local test constraint on the non-minimal f(R) model. Generally, we obtain the constraint $\psi_g-1<10^{-10}$. in the logarithmic coupling model, this constraint forces us to set the coupling parameter $\alpha$ to $10^{-10}$, which leads a small oscillating period $T\sim10^{-4}$, seeing Fig. 1.  However, its oscillating frequency is decreasing as the redshift increasing and the amplitude is not too large comparing with the normal f(R) theories. So, the singularity problem in normal f(R) model is no longer so serious when we consider the non-minimal coupling.
\section*{Acknowledgments}
The work was in part supported by NSFC Grant No. 10975005. We gratefully acknowledge Bin Chen for a careful reading
of the manuscript and insightful suggestions.

\end{document}